\documentclass[twocolumn,aps,floatfix,superscriptaddress]{revtex4-2}
\usepackage[utf8]{inputenc}

\usepackage{graphicx,color}
\usepackage{dcolumn}
\usepackage{cancel}
\usepackage{epsfig}
\usepackage{euscript}
\usepackage{amssymb}
\usepackage{amsmath}
\usepackage{amsthm}
\usepackage{newlfont}
\usepackage{bm,mathrsfs}
\usepackage{subfigure}
\usepackage{changes,todonotes}
\usepackage{subfigure}
\usepackage[colorlinks,linkcolor=blue,urlcolor=blue,citecolor=blue]{hyperref}
\usepackage[bbgreekl]{mathbbol}
\newcommand{\opunit}{\textrm{1}\kern-0.22em\textrm{l}}

\usepackage{makeidx}
\usepackage{comment}
\usepackage{soul}

\def\bea{\begin{eqnarray}}
\def\eea{\end{eqnarray}}
\def\ba{\begin{array}}
\def\ea{\end{array}}

\def\bea{\begin{eqnarray}}
\def\eea{\end{eqnarray}}
\def\ba{\begin{array}}
\def\ea{\end{array}}

\def\la{\langle}
\def\ra{\rangle}

\definecolor{dgreen}{rgb}{0,0.7,0}

\begin{document}

\title{Exact Volterra series for non-equilibrium mean field dynamics}

\author{Ion Santra}
\affiliation{Institut f\"ur Theoretische Physik, Universit\"at G\"ottingen }
\affiliation{Department of Physics and Astronomy, KU Leuven }
\author{Matthias Kr\"uger}
\affiliation{Institut f\"ur Theoretische Physik, Universit\"at G\"ottingen }

\begin{abstract}
	We derive a Volterra series for the  mean field dynamics of a driven interacting system, expressing the Volterra kernels in terms of response functions.  A gauge freedom  allows for different formulations. One gauge takes the form of dynamical density functional theory or model B dynamics, with forces deriving from a functional of the {\it  history} of density. The corresponding time nonlocal direct correlation functions obey time nonlocal Ornstein Zernike relations, with the familiar relations recovered in the zero frequency limit. Another gauge identifies an exact Rayleigh dissipation functional in terms of  response functions. These results allow for systematic improvement of established mean field formalisms. We exemplify these findings in simulations of driven interacting Brownian particles.
\end{abstract}

\maketitle
Describing driven systems far from equilibrium remains a formidable challenge, with a variety of recent developments, including non-equilibrium response theory~\cite{agarwal1972fluctuation,harada2005equality,speck2006restoring,nakamura2008fluctuation,baiesi2009nonequilibrium,kruger09,Basu2015,Basu18,maes2020response, caspers2025panoscopic}, fluctuating hydrodynamics~\cite{spohn2014nonlinear,derrida2019large,derrida2019large2}, large deviation theory~\cite{touchette2009large}, mode-coupling theory~\cite{reichman2005mode,brader2009glass,fuchs2009mode}, projection operators~\cite{zwanzig1973nonlinear,espanol2009,schilling2022coarse,ayaz2022generalized}, power functional theory~\cite{pft1,superadiabatic1,superadiabatic3}, stochastic field theories~\cite{dean1996langevin,illien2025}, or nonlinear Langevin approaches \cite{zwanzig1973nonlinear, kruger2016modified, CaspersLangevin}.

In many cases, the behavior of mean fields far from equilibrium is of interest, as e.g., in determination of non-equilibrium phases, structures, or  transitions~\cite{cross1993pattern,tauber2014critical}. There exist a variety of successful approaches for mean field dynamics, such as dynamical density functional theory (DDFT)~\cite{marconi1999,marconi2000dynamic,archer2004,chan2005time,espanol2009},  dynamical Curie-Weiss theory~\cite{chakrabarti1999dynamic}, dynamical Landau Ginzburg theory~\cite{hohenberg1977} or the Cahn-Hilliard equation~\cite{cahn1958free}. 
 These typically rely  on a  free-energy functional~\cite{munakata1989dynamical,dieterich1990nonlinear,archer2004,tarazona2014,te2020classical} to derive driving forces, thereby building on powerful equilibrium frameworks such as gradient expansion \cite{kardar2007statistical} or density functional theory (DFT) \cite{lebowitz1963statistical,evans1979,hansen2013, singh1991density,evans1992density,lowen2002density}.

These frameworks have been successfully applied to a broad range of systems, including colloids~\cite{penna2003dynamic,royall2007nonequilibrium,brader2011density,stopper2015modeling}, binary mixtures~\cite{onuki2002phase,goddard2013multi,stierle2021hydrodynamic}, glasses~\cite{fuchizaki2002dynamical,szamel2022alternative} or active particles~\cite{menzel2013traveling,menzel2016dynamical}. 
Recent developments include DDFT for shearing \cite{brader2011density,aerov14}, 
superadiabatic DDFT \cite{tschopp2022first,tschopp2024superadiabatic}, and identification of mobility kernels in model B dynamics \cite{akaberian2023nonequilibrium, thewes2024mobility}. Extensions to active model B~\cite{cates2015motility} and non-reciprocal Cahn-Hilliard equations~\cite{saha2020} have been shown  to describe motility induced phase separation and dynamical patterns in the context of active particles. There is, however, an apparent lack of exact closed form equations for the dynamics of a mean field. \textcolor{black}{As a consequence, it has remained open how these frameworks can be improved systematically from first principles.
 Similarly, it is unknown how the dissipation kernel of a power functional theory (PFT) can be systematically constructed.}

In this manuscript, we derive an exact Volterra expansion for the dynamics of a mean field, driven away from equilibrium by a time dependent external potential. 
Using nonlinear response theory, we find hierarchical relations for the Volterra kernels in terms of response functions. We emphasize an inherent freedom in this expansion, and we identify two useful choices. One choice yields a form in close similarity to dynamical density functional theory or model B dynamics, but with interactions encoded in a causal density-\emph{history} functional. The corresponding time nonlocal direct correlation functions are shown to obey time non-local Ornstein Zernike relations, reducing to the known equilibrium ones at zero frequency. 
Another gauge yields an exact -- in a series sense -- expression for the time nonlocal Rayleigh dissipation functional in an Onsager formulation or power functional theory, in terms of response functions.  
These developments allow for a systematic improvement of existing mean field theories such as dynamical density functional theory, Landau Ginzburg theory, power functional theory, and others, and are expected to be especially beneficial in cases of slow dynamics with broad distributions of time scales\textcolor{black}{, where current approaches are imprecise. We illustrate  this in an example of Brownian dynamics simulations of an interacting fluid.}

Consider a system of degrees of freedom $X$ governed by the Hamiltonian 
\begin{align}
\mathcal  H_f(X,t)=\mathcal H_0(X)+\int dx\, \hat\rho_{x}(X)\,f_{x,t}.
\end{align}
 The second term on the right hand side drives the system away from a reference equilibrium according to $\mathcal H_0(X)$, temperature $T$ and chemical potential $\mu$ (we use units with $k_BT=1$ throughout). $\hat\rho_x(X)$ is a phase space function, allowed to  explicitly depend on spatial coordinate $x$, with $x\in \mathbb{R}^d$   \footnote{In $d$-dimensions, $\int dx$ denotes a $d$-dimensional integral over space. For  ease of notation, we suppress vector notation for $x$ throughout.}. $\hat\rho_x(X)$ is thus a field observable, e.g., the microscopic density defined in Eq.~\eqref{density:op} below; $f_{x,t}$ is a deterministic perturbation in form of a potential, which may depend on  $x$ and time $t$. 
In this perturbed system, the field takes the mean  
$\rho_{x,t}\equiv\la \hat\rho_{x}\ra_f$, a function of $x$ and $t$, the latter dependence entering via the non-equilibrium phase space distribution used in $\langle \dots \rangle_f$.  \textcolor{black}{The goal of this work is to find a closed evolution equation for the mean field $\rho_{x,t}$}. We write $\rho^{(0)}_x\equiv\la \hat\rho_{x}\ra_0$, the time independent mean field of the equilibrium system. 

The dynamics of the nonequilibrium system is defined by the response, expanded in powers of $f$ via the following Volterra series (throughout, $z=\{x,t\}$ denotes a combined space-time coordinate, with integrals implying $\int dz_i=\int_{-\infty}^t dt_i\int dx_i$), and we introduce $\phi_z\equiv \rho_z-\rho_x^{(0)}$
\begin{align}
\phi
=
\sum_{n=1}^{\infty}
\frac{1}{n!}\mathbb{\Gamma}^{(n)}\left[f,\ldots,f\right].
\label{eq:response}
\end{align}
Eq.~\eqref{eq:response} uses a compact notation in terms of the multilinear operator
 $\mathbb{\Gamma}^{(n)}$, with $\mathbb{\Gamma}^{(n)}\left[f,\ldots,f\right]\equiv \int \dots \int 
 \Gamma^{(n)}_{z,z_1, \dots,z_n} \prod_{i=1}^nf_{z_i} dz_i$, with  the response function of order $n$ (with $t_i\leq t$)
\begin{align}
\Gamma^{(n)}_{z,z_1,\cdots, z_n}=\frac{\delta^n \rho_{z}}{\delta f_{z_1}\cdots\delta f_{z_n}}\Bigg|_{f=0}.
\label{def:response}
\end{align}
We seek a closed evolution equation for the mean field. As it results from eliminating microscopic degrees of freedom, we expect it to be nonlocal in space and time, 
\begin{align}
\partial_t\phi &=\beta[\rho]+\bbchi[\rho] f.
\label{eq:ddft1}
\end{align}
$\beta[\rho]=\sum_{n=1}^{\infty}\bbbeta^{(n)}[\phi\dotsc\phi]$ is a path-functional of the mean field, \textcolor{black}{which is function valued, i.e., which depends on $x$ and $t$.} $\bbchi[\rho]=\sum_{n=0}^{\infty}\bbchi^{(n)}[\phi\dotsc\phi]$ is an operator valued functional, \textcolor{black}{i.e., dependent on two arguments,}  acting on the external field, \textcolor{black}{i.e., $(\bbchi f)_z=\int dz'\chi_{z,z'}f_{z'}$}. 
Eq.~\eqref{eq:ddft1} is thus a Volterra expansion in powers of the mean field, where kernels $\bbbeta^{(n)}$ and $\bbchi^{(n)}$ are nonlocal in space and time, see App.~\ref{app1} for the explicit version.

 The form of Eq.~\eqref{eq:ddft1} is ubiquitous and appears frequently in coarse grained descriptions of interacting systems. The contribution of this manuscript is the exact identification of the  kernels appearing in Eq.~\eqref{eq:ddft1}, order by order, in terms of the response functions in Eq.~\eqref{eq:response}. 
 To this end, we take functional derivatives of Eq.~\eqref{eq:ddft1} with respect to  $f$, evaluated at $f=0$.  
 The differentiated fields $\rho$ are thereby replaced by the response functions using Eq.~\eqref{def:response}. 
 This yields a hierarchy of relations, of which the first two are,
\begin{subequations}\label{eq:recur}
\begin{align}
&\partial_t\mathbb{\Gamma}^{(1)}=\bbchi^{(0)}+\bbbeta^{(1)}\mathbb{\Gamma}^{(1)} ,\label{eq:1storder1}\\
&\partial_ t\mathbb\Gamma^{(2)}=2\bbchi^{(1)}\mathbb\Gamma^{(1)}+\bbbeta^{(1)}\mathbb\Gamma^{(2)}+2\bbbeta^{(2)}\mathbb\Gamma^{(1)}\mathbb{\Gamma}^{(1)}.\label{eq5b}
\end{align}
\end{subequations}
See App.~\ref{app:l1} for the relation at arbitrary order. $\mathbb{\Gamma}^{(1)}$,  $\bbchi^{(0)}$ and $\bbbeta^{(1)}$ are operators (kernels) with two space-time arguments, and their product implies integration over a joint space and time argument, \textcolor{black}{e.g., $(\bbbeta^{(1)}\mathbb{\Gamma}^{(1)})_{z,z'}=\int dz'' \beta^{(1)}_{z,z''}\mathbb{\Gamma}^{(1)}_{z'',z'}$}. $\mathbb{\Gamma}^{(2)}$,  $\bbchi^{(1)}$ and $\bbbeta^{(2)}$ carry, accordingly, three arguments, and $\bbbeta^{(2)}\mathbb\Gamma^{(1)}\mathbb{\Gamma}^{(1)}$ implies contracting each one argument of $\bbbeta^{(2)}$ with ${\Gamma}^{(1)}$,  
see App.~\ref{app1} for explicit integral forms.

Eqs.~\eqref{eq:recur} demonstrate that $\bbchi^{(n-1)}$  and $\bbbeta^{(n)}$ are determined in terms of the response functions $\mathbb{\Gamma}^{(i)}$ with $i\leq n$. In other words, the Volterra kernels of a dynamical mean
field theory can be expressed in terms of non-linear re-
sponse kernels. This is the first main result of this work which has a wide range of applicability,
including dynamics of conserved or non-conserved fields.

Notably, Eqs.~\eqref{eq:recur} can only be solved for $\beta$ in terms of $\chi$, or vice versa~\footnote{
For example,
\textcolor{black}{$\bbchi^{(0)}\to\bbchi^{(0)}+\mathbb{A}$ and $\bbbeta^{(1)}\to\bbbeta^{(1)}-\mathbb{A}(\mathbb{\Gamma}^{(1)})^{-1}$}
leave Eq.~(5a) invariant.}. \textcolor{black}{In absence of further constraints, there is a freedom and we explore two specific choices in the following.} 
 We therefore consider the special case of a  single-component fluid and  the microscopic density \cite{hansen2013} \begin{align}
\hat{\rho}_x=\sum_{i=1}^N\delta_{x-x_i},\label{density:op}
\end{align}
with $x_i$ the position (vector) of particle $i$.  $\rho_x^{(0)}$ is the equilibrium density profile.

The first choice uses $\bbchi$ local in space and time, i.e., 
$\bbchi_{\mathrm{D}\,z,z'}=\delta_{z-z'}\nu\nabla\cdot\rho_z \nabla$, with a bare mobility $\nu$ \footnote{Index 'D' marks the 'DDFT-gauge' introduced below.}\footnote{The temporal delta distribution is understood as $\delta_{t-t'-0^+}$.}. Splitting  the first term of Eq.~\eqref{eq:ddft1} into a product of $\bbchi$ and an internal force, this yields a form similar to standard DDFT \cite{archer2004}, 
\begin{align}
\partial_t\phi_z=\nu\nabla\cdot\rho_z\nabla\left[\mathcal U_z[\rho]+f_z\right].
\label{eq:ddft}
\end{align}
$\mathcal U$ is a function-valued path-functional of the mean density, i.e., $\mathcal U_{x,t}$ depends on the spatiotemporal evolution of $\rho_{t'}$, with $t'\leq t$. Replacing $\mathcal U_{x,t}[\rho]$ by $\frac{\delta\mathcal F}{\delta\rho_x}$, i.e., the derivative of the equilibrium free energy $\mathcal F$, 
recovers standard DDFT \cite{archer2004}.
How is the exact form of $\mathcal U$ found? We expand it as 
\begin{align}
    \mathcal U[\rho]=\mathcal U[\rho^{(0)}]+ \sum_{n=1}^{\infty}\left[
-\frac{1}{n}\left(\frac{-\phi}{\rho^{(0)}}\right)^n 
-\frac{\mathbb C^{(n+1)}\,
[\phi,\ldots,\phi]}{n!}\right]
\label{eq:defU}
\end{align} 
with the first term in the sum resulting from the ideal gas contribution.

We thus need to find the 
 time non-local direct correlation functions $\mathbb{C}^{(n)}$ that build the expansion coefficients of   $\mathcal U$,
 as equilibrium direct correlations \textcolor{black}{build the expansion coefficients of the equilibrium analog}~\cite{hansen2013}.  
The $\mathbb{C}^{(n)}$ are found using the algorithm outlined in Eqs.~\eqref{eq:recur}, and they are  expressed in terms of the response kernels $\mathbb{\Gamma}^{(n)}$ \footnote{Rewriting Eq.~\eqref{eq:ddft} as $\partial_t\phi\equiv\partial_t\Gamma[f]=\bbchi_\text{D}[\rho]\left(\mathcal U[\rho]+\Gamma^{-1}[\phi]\right)$ shows that the  coefficients $\mathbb{C}^{(n)}$ are expressed in terms of the response $\Gamma$.}. 
We provide explicitly the leading coefficient, $\mathbb{C}^{(2)}$, which reads (with Laplace transform, $q_s=\int_0^\infty dt e^{-st}q_t$),
 \begin{align} C^{(2)}_{s,x,x'}
 =\frac{\delta_{x-x'}}{\rho^{(0)}_x}+\frac{1}{\Gamma^{(1)}_{s,x,x'}}-\frac{s}{\chi^{(0)}_{\mathrm{D}\,x,x'}}, \label{eq:dynamicOZ2}
\end{align}
where inverses are interpreted as  operator inverses. 
Using Eq.~\eqref{eq:dynamicOZ2}, $\mathcal U$ in Eq.~\eqref{eq:ddft}  
is given to leading order in $\phi$ by $\mathcal U-\mathcal U[\rho^{(0)}]=\phi_{z}/{\rho^{(0)}_x}-\int dz_1 C^{(2)}_{z,z_1}\phi_{z_1} +\mathcal{O}( \phi^2)$,
making explicit that $\mathcal U$ is indeed a path-functional of density.

The time nonlocal Ornstein Zernike relation in Eq.~\eqref{eq:dynamicOZ2} shows intriguing similarity to the one  familiar from equilibrium. Indeed, sending $s\to0$ in Eq.~\eqref{eq:dynamicOZ2}  yields
\begin{align}
 C^{(2)}_{s=0,x,x'}
 &=\frac{\delta_{x-x'}}{\rho^{(0)}_x}+\frac{1}{\Gamma^{(1)}_{s=0,x,x'}}
 =c^{(2)}_{x,x'},
 \label{eq:genOZstatic}
\end{align}
which is the second order equilibrium Ornstein-Zernike (OZ) relation, expressing the direct correlation function $c^{(2)}_{x,x'}$ in terms of the static response function $\Gamma^{(1)}_{s=0}$ \cite{hansen2013}. In other words, the time nonlocal Ornstein Zernike relation of Eq.~\eqref{eq:dynamicOZ2} is found from the equilibrium one by replacing the static response function by the dynamic one, $\mathbb{\Gamma}^{(1)}_{s} (\bbchi_\mathrm{D}/s-\mathbb{\Gamma}^{(1)}_{s})^{-1}\bbchi_\mathrm{D}/s$. Higher order Ornstein-Zernike relations are given in App.~\ref{appOZ}.  

This remarkable relation between the time nonlocal and the equilibrium  direct correlation function can be   shown to higher orders  [see App.~\ref{app:general_sto0_limit}], i.e., we find
\begin{align}
C^{(n)}_{\{s_1, \dots, s_{n-1}\}=0}
=
c^{(n)},
\qquad n\geq2.
\label{eq:s=0}
\end{align}
Consequently, for a time-independent density profile,
all temporal integrations in  $\mathcal U$ select the zero-frequency coefficients, and $\mathcal U_x$ reduces to $\mathcal U_x=\delta\mathcal F/\delta\rho_x$, i.e., the derivative of the equilibrium free energy. Thus, as an important consistency check, for a time-independent external field $f_x$, the corresponding equilibrium density is a stationary solution of Eq.~\eqref{eq:ddft}.

A second useful gauge takes the following form \cite{thewes2024mobility}
\begin{align}
   \partial_t\phi_{z}
&=
\int d z'
\chi_{\mathrm{P}\,z,z'}\left[
\frac{\delta\mathcal{F}[\rho]}{\delta\rho_{z'}}+f_{z'}\right]\label{eq:ddft2}.
\end{align}
In Eq.~\eqref{eq:ddft2}, the \emph{equilibrium} free energy functional $\mathcal{F}$ appears, and the mobility kernel $\bbchi_\mathrm{P}$ 
is the object of interest to be determined \footnote{Index 'P' marks the `PFT-gauge'.}. Before doing so, we note that Eq.~\eqref{eq:ddft2} may be inferred from the Onsager principle \cite{Onsager1931a,Onsager1931b,Doi2011}, using the following Rayleigh- or power functional of mean density and mean current $\mathbf{J}$ \cite{pft1,treffenstadt2020}, obeying continuity  $\partial_t \phi=-\nabla \cdot \mathbf{J}$,
\begin{align}
\mathcal{R}_t[\mathbf{J},\rho]=&\int_{-\infty}^t dt'\int  dx\int dx'\mathbf{J}_z\cdot\nabla\chi_{\mathrm{P}\,z,z'}^{-1}[\rho]\nabla'\cdot\mathbf{J}_{z'}\notag\\&+\int dx\mathbf{J}_{z}\cdot\nabla\left[ \frac{\delta \mathcal{F}[\rho]}{\delta\rho_z}+f_z\right].\label{eq:PFT}
\end{align}
The first term in Eq.~\eqref{eq:PFT} corresponds to dissipation, while the second resembles the change of free energy \cite{pft1}. Notably, we find that 
the variational principle 
$    \frac{\delta \mathcal{R}_t}{\delta \mathbf{J}_z}=0$ with  continuity yields Eq.~\eqref{eq:ddft2}, making this gauge choice thus equivalent to power functional theory \cite{pft1}.  Importantly, the dissipation kernel $\bbchi_\mathrm{P}^{-1}$ appearing in $\mathcal{R}_t$ is generally considered unknown \cite{superadiabatic1,superadiabatic2}. Another main finding of this manuscript is thus the explicit, exact, form of  $\bbchi_\mathrm{P}^{-1}$ in terms of nonlinear response functions \footnote{Introducing the response $\phi=\Gamma[f]$ with inverse $\Gamma^{-1}$ and its static analog $\Gamma_0^{-1}$ yields  $\partial_t\Gamma[f]=\bbchi_\mathrm{P}(-{\Gamma}_0^{-1}[\phi]+{\Gamma}^{-1}[\phi])$. It may be inverted perturbatively, yielding the  result of Eq.~\eqref{eq:chi1}.}. To leading order in $\phi$ (see App.~\ref{app:chi} for higher orders)
\begin{align}
\frac{1}{\bbchi^{(0)}_{\mathrm{P}\,s} }&=\frac{1}{s\mathbb{\Gamma}^{(1)}_s}-\frac{1}{s\mathbb{\Gamma}^{(1)}_{s=0}}.\label{eq:chi-1}
\end{align}
Eq.~\eqref{eq:chi-1} gives the remarkable result that the dissipation kernel is given in terms of the difference of inverse of dynamic and static response. This yields the following setting for the power functional in Eq.~\eqref{eq:PFT}: The free energy in Eq.~\eqref{eq:PFT} is given in terms of \emph{static} response functions, while the dissipation kernel is given in terms of \emph{dynamic} response functions. Eq.~\eqref{eq:chi-1}  shows that $\bbchi_\mathrm{P}$,  the mobility kernel in  ``PFT-gauge" is nonlocal in space and time, while $\bbchi_\mathrm{D}$, the kernel in ``DDFT-gauge" defined above, is local.

A gauge freedom can emerge when mathematical objects are not accessible to experiment, like the vector potential in electrodynamics \cite{Jackson}. We argue that, here, the mobility kernel has in general no separate definition or measurability, giving rise to the mentioned freedom.

 For overdamped Brownian particles subject to  a bare mobility $\nu$ with respect to a resting background fluid \cite{dhont_introduction_1996}, the forces between  particles and fluid are well defined, and the DDFT gauge of  Eq.~\eqref{eq:ddft} is particularly meaningful. $\mathcal U$ in Eq.~\eqref{eq:ddft}  can then be  identified via $\rho_{x,t}\nabla\mathcal U_{x,t}[\rho]=\int dx'\rho^{(2)}_{x,x',t}\nabla_x u_{|x-x'|}$, up to a divergence-free part, with the two-body density $\rho^{(2)}$ and pairwise interaction potential $u$ \cite{archer2004}.

Eq.~\eqref{eq:ddft2} is appealing because it uses the driving forces resulting from the equilibrium free energy. 
Applications thus include linear stability analysis and early time dynamics of unstable modes \cite{archer2004}. Eq.~\eqref{eq:ddft2} is also conceptually important as it  transforms into a Rayleigh- or power functional.

\begin{figure*}
    \centering
    \includegraphics[width=\linewidth]{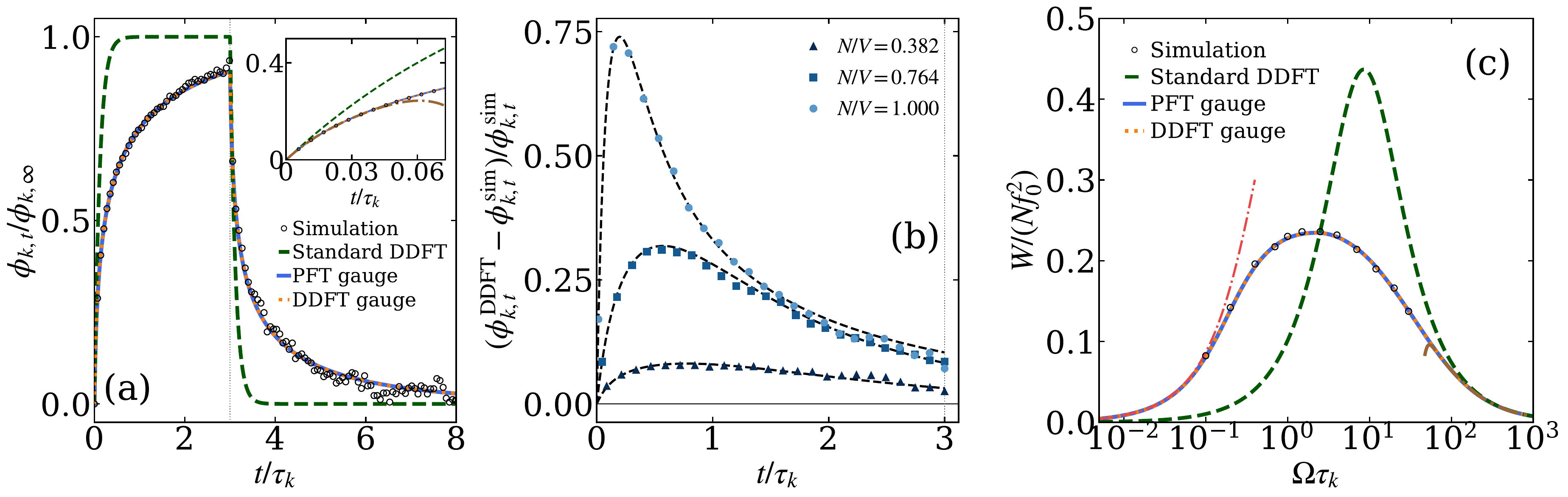}
    \caption{Mean field dynamics and work done for overdamped Brownian particles of diameters $\sigma$ interacting with a WCA potential, from simulations (data points) and different theories (lines as labeled). $k\sigma=6.89$ in all panels.
    Panel (a): Evolution of the mean density  following the perturbation of  Eq.~\eqref{eq:perturb1} for $N/V=1$ and $t_0=3\tau_k$. Inset: Zoom of short time behavior, with brown dashed line depicting the result of the large-$s$ expansion of $C^{(2)}_{k,s}$ in Eq.~\eqref{eq:dynamicOZ2}. Panel (b):  Difference between simulations and standard DDFT  (data points)  for different values of $N/V$, with dashed lines  from  Eq.~\eqref{eq:ddft} or \eqref{eq:PFT}.  Panel (c): Work per cycle for a time periodic potential as a function of frequency $\Omega$ for $N/V=1$. Dash-dotted lines show the leading corrections to DDFT obtained from the small-$s$ (red) and large-$s$ (brown) expansions of $C^{(2)}_{k,s}$,  respectively. 
    }
\label{f:harmonic-chain}   
\end{figure*}
 To illustrate the validity of Eqs.~\eqref{eq:ddft} and \eqref{eq:ddft2}, we perform simulations on $N$ overdamped Brownian particles in a three dimensional box of volume $V$ interacting via a Weeks-Chandler-Anderson potential, 
  see SM \cite{SM} for details. We choose the perturbation 
 \begin{align}
 f_{\mathbf{x},t} = -f_0\cos(\mathbf{k}\cdot\mathbf{x})\,\Theta(t)\Theta(t_0-t),\label{eq:perturb1}
 \end{align}
 i.e., a potential periodic in space, switched on at time $t=0$ and switched off at time $t=t_0$.
 We extract the response $\Gamma^{(1)}_{k,t}$ from  simulations, see \cite{SM},  
 noting that the linear response is sufficient for the values $f_0=0.1/S_k$ chosen, with $S_k$ the static structure factor.
 Eq.~\eqref{eq:dynamicOZ2} is used to determine the
time-nonlocal direct correlation function $\mathbb C^{(2)}_{k,t}$ (DDFT-
gauge), and Eq.~\eqref{eq:chi-1} is used to determine the  mobility kernel
$\bbchi^{(0)}_{ \text{P}\,k,t}$ (PFT-gauge). These  are monotonically decaying functions of time, see SM \cite{SM}, and are thus expected to be amenable to  approximations in future work. Using these in Eqs.~\eqref{eq:ddft}, and \eqref{eq:ddft2}, respectively, yields the non-equilibrium mean field $\phi_{k,t}$, which is compared to the outcome of our simulations. For comparison, we also show the prediction of standard DDFT, for which the needed static correlation function $c_k^{(2)}$ is extracted from  simulations. Notably, for small $f_0$, DDFT predicts a single exponential
relaxation with relaxation time  $S_k/(\nu k^2)$
, exact at short
times~\cite{fuchs2009mode, kruger_gaussian_2017}. 

Figure~\ref{f:harmonic-chain}(a) shows the density evolution for $N/V=1$ \footnote{$N/V=1$ lies above the freezing density of the WCA fluid at $N/V \approx 0.964$ for the chosen parameters~\cite{SM,khrapak2024freezing}, which we however do not observe in the finite size system.} as a function of time, scaled by  the relaxation time $\tau_k$\footnote{The relaxation time is defined as $\tau_k=\int_0^\infty dt\,t\,\Gamma^{(1)}_{k,t}/\int_0^\infty dt\,\Gamma^{(1)}_{k,t}$.}. 
Eqs.~\eqref{eq:ddft} and \eqref{eq:PFT} both very precisely predict the outcome of our simulations, while, as expected,  DDFT  predicts too fast relaxation. The inset shows how DDFT can be systematically improved by displaying the   prediction of Eq.~\eqref{eq:ddft} when replacing $\mathbb{C}^{(2)}$ by its large $s$ expansion; it yields a precise result for short times.

Figure~\ref{f:harmonic-chain} (b) shows the deviation of DDFT from simulations (data points) and Eqs.~\eqref{eq:ddft} and \eqref{eq:ddft2} (dashed lines) for different densities. The latter match the simulations very precisely. The deviations increase with density, indicating that the here developed theory is especially important and needed at high densities with slow dynamics.


This analysis has direct thermodynamic implications, as seen from the  average work $W$ done by the external field during a time interval $[0,2\pi/\Omega]$,
$W=\int_0^{2\pi \Omega^{-1}}dt\int_V dx\, \rho_{x,t}\,\partial_t f_{x,t}$,
requiring an accurate determination of the time dependent mean density. 
Figure~\ref{f:harmonic-chain}(c) shows $W$  for the  system of Fig.~\ref{f:harmonic-chain} (a), using a time periodic potential, $f_{\mathbf{x},t} = f_0\cos(\mathbf{k}\cdot\mathbf{x})\cos(\Omega t)$, as a function of $\Omega$. We find that the mean fields from  Eqs.~\eqref{eq:ddft} and \eqref{eq:PFT} precisely predict $W$ from simulations, while standard DDFT does so  only for large $\Omega$. The difference between $W$ and its DDFT-prediction is extensive (see App.~\ref{app:work}), i.e., it remains finite in the thermodynamic limit.

The presented formalism allows systematic improvements over standard DDFT for both small and large driving frequencies. For small $\Omega$, the red (dash-dot) curve in Fig.~\ref{f:harmonic-chain} (c) shows the result of expanding $\mathbb{C}^{(2)}$  around $s=0$, i.e., using $ C^{(2)}_{k,s}
\approx c_k^{(2)}
+s\,\partial_s C^{(2)}_{k,s}|_{s=0}$ in Eq.~\eqref{eq:defU}. For large $\Omega$, the large $s$ expansion of $ C^{(2)}_{k,s}$ yields the Brown line in panel (c), compare also the inset in panel (a).


We developed an exact Volterra series expansion for a mean field of  a system driven by an external potential, with  coefficients  given in terms of response functions. 
 We demonstrated that this formalism allows for systematic improvements over existing mean field theories, which are highly needed for precise predictions. We highlight different gauges: In a DDFT gauge, the equilibrium free energy landscape is promoted to a history-dependent functional, with the static response functions replaced by the dynamic ones in the corresponding Ornstein Zernike relations. In a PFT gauge, we obtain the exact perturbative expansion of the dissipation kernel and power functional.
 While we highlight simple cases, this formalism is applicable to myriad situations, e.g., mixtures, complex fluids,  glassy materials, in bulk or in confinement. It can equivalently be used to systematically  improve other mean field approaches such as Landau Ginzburg Theory, or the Rayleigh dissipation functional.

Future work will extend this formalism to biological or active fluids, and to other types of driving, such as thermal quenches, using recent advances in non-equilibrium response theory~\cite{Basu18,basu2015nonequilibrium}. It will also be important to  explore closure relations or other approximations in comparison to numerical and experimental data.

\emph{Acknowledgements} The authors thank Daniel de la Heras and Martin Oettel for useful suggestions on the gauge choices.
IS acknowledges funding from the European Union’s Horizon Europe 2024 research and innovation programme under the Marie Sklodowska-Curie (HORIZON-TMA-MSCA-PF-EF) grant agreement No. 101205210. MK acknowledges funding from the German Research Foundation (DFG) under grant number KR3844/5-1. 

\bibliography{ref}
\onecolumngrid

\newpage
\bigskip \bigskip 
\centerline {\bf End Matter}
\bigskip

\twocolumngrid

\appendix
\section{Explicit notations}\label{app1}
In this section, we provide, for completeness, the explicit integral forms of the operator equations in the main text. For example, Eq.~\eqref{eq:ddft1}, written explicitly to
second order in $\phi$, is
\begin{align}
    \partial_t\phi_z&=\int dz' \beta^{(1)}_{z,z'}\,\phi_{z'}+\int dz'dz''\beta^{(2)}_{z,z',z''}\phi_{z'}\phi_{z''}+\cdots\nonumber\\
    &+\int dz'\,\chi^{(0)}_{z,z'}f_{z'}
+\int dz'\,dz''\,\chi^{(1)}_{z,z',z''}\phi_{z'} f_{z''}
+\cdots.
\end{align}
More generally, the $n$th order term
\begin{align}
    \bbbeta^{(n)}[\phi,\ldots,\phi]&=\int dz_1\cdots \int dz_n \beta^{(n)}_{z,z_1,\cdots,z_n}\prod_{i=1}^n \phi_{z_i}\\
\bbchi^{(n)}[\phi,\ldots,\phi]f&=  \int dz_1\cdots \int dz_n\int dz' \bbchi^{(n)}_{z,z_1,\cdots,z_n}\nonumber\\
&\times\left(\prod_{i=1}^n \phi_{z_i}\right)  f_z'
\end{align}

The hierarchy of Eqs.~\eqref{eq:recur} therefore read,
\begin{align}
&\partial_t\Gamma^{(1)}_{z,z'}
=
\chi^{(0)}_{z,z'}
+
\int dz_1\,
\beta^{(1)}_{z,z_1}\Gamma^{(1)}_{z_1,z'}\\
&\partial_t\Gamma^{(2)}_{z,z',z''}
=
\int dz_1\,
\chi^{(1)}_{z,z_1,z'}\Gamma^{(1)}_{z_1,z''}
+
\int dz_1\,
\chi^{(1)}_{z,z_1,z''}\Gamma^{(1)}_{z_1,z'}
\nonumber\\
&~~~~~+
\int dz_1\,
\beta^{(1)}_{z,z_1}\Gamma^{(2)}_{z_1,z',z''}
\nonumber\\
&~~~~~+
\int dz_1\,dz_2\,
\beta^{(2)}_{z,z_1,z_2}
\Big[
\Gamma^{(1)}_{z_1,z'}\Gamma^{(1)}_{z_2,z''}
+
\Gamma^{(1)}_{z_1,z''}\Gamma^{(1)}_{z_2,z'}
\Big].
\end{align}

\section{General expression for the kernels}
\label{app:l1}
In this section, we derive the general relation connecting the
kernels $\bbbeta^{(n)}$ and $\bbchi^{(n)}$ of the mean-field equation
\begin{equation}
\partial_t\rho=\bbbeta[\rho]+\bbchi[\rho]f
 \label{S_mean_field}
\end{equation}
to the response functions $\mathbb{\Gamma}^{(n)}$, defined by
\begin{equation}
 \phi =
 \sum_{n=1}^{\infty}
 \frac{1}{n!}\,
 \mathbb{\Gamma}^{(n)}[f,\ldots,f].
 \label{S_response}
\end{equation} 
Taking $m$ functional derivatives of Eq.~\eqref{S_mean_field}
with respect to $f$ and evaluating at $f=0$ gives
\begin{align}
\partial_t \mathbb\Gamma^{(m)}&=\sum_{j=1}^{m}  \sum_{w=0}^{m-1}w!
\sum_{\mathcal P\in\mathcal P(I_m\setminus\{j\},w)}
\bbchi^{(w)}\left[\mathbb{\Gamma}^{(|B_1|)},\dotsc,\mathbb{\Gamma}^{(|B_w|)}\right]
\nonumber\\
&+
\sum_{j=1}^{m}j!\sum_{\mathcal P\in\mathcal P(I_m,j)}
\bbbeta^{(j)}\left[
\mathbb{\Gamma}^{(|B_1|)},\dotsc,\mathbb{\Gamma}^{(|B_j|)}\right].
\label{A1}
\end{align}
where $I_m=\{1,2,\dots,m\}$, and $\mathcal P(I_m,w)$ denotes the set of all partitions of $I_m$ into $w$ nonempty, disjoint blocks $ B_\ell$.
The first sum collects contributions from the mobility kernels $\bbchi^{(w)}$, while the second sum involves the drift kernels $\bbbeta^{(j)}$. Putting $m=1,\,2$ leads to Eq.~\eqref{eq:recur} in the main text. It is clear from the general expression, that at order $m$, the only new operators are $\bbchi^{(m-1)}$ and $\mathbb\Gamma^{(m)}$, all the other terms involve lower order operators.

\section{Higher order time nonlocal OZ relation}\label{appOZ}
In this section, we give explicitly the third order Ornstein Zernike relation, demonstrating that it is expressed in terms of response functions up to order two. In the DDFT gauge, $\beta[\rho]=\bbchi_\text{D}[\rho]\mathcal U[\rho]$, implying that,
\begin{align}
    \bbbeta^{(1)}&=\bbchi_\text{D}^{(0)}\left( \mathbb{I}^{(1)}-\mathbb C^{(2)}\right),\\
    \bbbeta^{(2)}&=\frac12 \bbchi^{(0)}_\text{D}\left(-\mathbb{I}^{(2)} -\mathbb C^{(3)}\right)+\bbchi_\text{D} ^{(1)}\left( \mathbb{I}^{(1)}-\mathbb C^{(2)}\right)
\end{align}
where $ \mathbb{I}^{(1)}\equiv\delta_{z-z_1}/\rho_x$, and $\mathbb{I}^{(2)}\equiv\delta_{z-z_1}\delta_{z-z_2}/\rho_x^2$ denote two and three point kernels. Using these in Eq.~\eqref{eq5b}, we get,
\begin{align}
   \mathbb C^{(3)}&=-\mathbb{I}^{(2)}-\left(\bbchi_{ \text D}^{(0)}\right)^{-1}\Bigg( \partial_t\mathbb \Gamma^{(2)}-2\bbchi_\text{D}^{(1)}\mathbb\Gamma^{(1)} \nonumber\\
&\quad-\bbchi_\text{D}^{(0)}\left(\mathbb I^{(1)}-\mathbb C^{(2)}\right)\mathbb \Gamma^{(2)}\Bigg)(\mathbb\Gamma^{(1)})^{-1}(\mathbb\Gamma^{(1)})^{-1}\nonumber\\
&\quad +2\left(\bbchi_{\text D}^{(0)}\right)^{-1}
\bbchi_{\text D}^{(1)}
\left(\mathbb I^{(1)}-\mathbb C^{(2)}\right).
\end{align}
Using the equilibrium static response functions, along with Eq.~\eqref{eq:genOZstatic}, it can be verified explicitly that the above form reduces to the third order equilibrium Ornstein-Zernike relation in the limit of $\{s_i\}\to0$.

\section{Proof of general static limit of $C^{(n)}$}\label{app:general_sto0_limit}
Are the $C^{(n)}$ with $n>2$ linked to the corresponding equilibrium direct correlation functions as well? To this end, let us be reminded that $\mathcal{U}[\phi]$ in Eq.~\eqref{eq:ddft} is a functional of $\phi_z$ in space and time. It is beneficial to evaluate $\mathcal{U}$ for the special case of a time independent density, i.e., for the {\it test field} $\phi_{x,t}=\Phi_{x}$ for all $t$. For such a density field, all the temporal integrations in the expansion of $\mathcal{U}$ selects the corresponding kernels  $C^{(n)}$ at $\{s_i\}=0$. Moreover, $C^{(n)}$s are Taylor expansion coefficients of $\mathcal U$ (apart from its ideal-gas part, see Eq.~\eqref{eq:defU}), i.e., successive functional derivatives of the same functional. Since the expansion holds for an arbitrary reference profile, they obey the recursion relation
\begin{align}
  C^{(n+1)}_{x_1,\cdots,x_n,\{s_i\}=0}=\frac{\delta   C^{(n)}_{x_1,x_2,\cdots,x_{n-1},\{s_i\}=0}}{\delta \rho_{ x_n}}.\label{eq:recursion}
\end{align} 

In Eq.~\eqref{eq:dynamicOZ2} in the main text, we showed that the second order kernel $C^{(2)}$ in the $s\to 0$ limit satisfies equilibrium OZ relation. It is worth pointing that, the reference state is arbitrary, and Eq.~\eqref{eq:dynamicOZ2} holds for any equilibrium profile, so that so that $\mathbb C^{(2)}_{s=0}[\rho^{(0)}]
=c^{(2)}[\rho^{(0)}]$ is a functional identity, and can be differentiated with respect to the density.
Realizing that the equilibrium direct correlation functions obey the recursion in Eq.~\eqref{eq:recursion}, we have thus proven by induction that for $n\geq2$
\begin{align}
    \bar C^{(n)}_{\{s_i\}=0}= c^{(n)}\label{eq:s=0}
\end{align}
with $c^{(n)}$ the equilibrium direct correlation function of order $n$.

Consequently, for a time-independent density profile,
all temporal integrations in the expansion of $\mathcal U$
select the equilibrium direct correlation functions, and the
time-nonlocal force functional reduces to that derived from the equilibrium free-energy
functional,
\begin{align}
 \mathcal U_x[\rho^{(0)}+\Phi]
    =
    \frac{\delta\mathcal \mathcal F}{\delta\rho_x}\bigg|_{\rho^{(0)}+\Phi},    \label{free-energy-limit}
\end{align}
Eq.~\eqref{free-energy-limit} provides an important cross check of our formalism. 
It demonstrates that, for the case of a time independent external field $f_x$, 
he equilibrium density profile, determined by $\delta\mathcal F/\delta\rho_x+f_x=\mu$] is a stationary solution, independent of the gauge choice.
\section{Higher orders of dissipation kernel}\label{app:chi}
In this section, we present the PFT gauge mobility, i.e., the corresponding dissipation kernel beyond the leading order given in Eq.~\eqref{eq:chi-1}. The  mobility is expanded as $\bbchi_{\text P}[\rho]
=
\bbchi_{\text P}^{(0)}
+
\bbchi_{\text P}^{(1)}[\phi]
+\mathcal O(\phi^2)$, yielding the dissipation kernel, i.e., its inverse,
\begin{align}
    \chi_\text{P}^{-1}\psi&=(\bbchi_\text{P}^{(0)})^{-1}\psi-(\bbchi_\text{P}^{(0)})^{-1}\bbchi_\text{P}^{(1)}[\phi,(\bbchi_\text{P}^{(0)})^{-1}\psi]+\mathcal O(\phi^2).\label{chipapp}
\end{align}
$\bbchi_P^{(1)}$ is obtained from demanding that Eq.~\eqref{eq:ddft2} yields the correct dynamics to the given order. This  
expresses $\bbchi_P^{(1)}$ in terms of direct correlation functions,
\begin{align}
    2&\bbchi_{\text{P}}^{(1)}\mathbb{\Gamma}^{(1)}
+
2\bbchi_{\text{P}}^{(1)}
\left(\mathbb I^{(1)}-c^{(2)}\right)
\mathbb{\Gamma}^{(1)}\mathbb{\Gamma}^{(1)}
\nonumber\\
&\quad=
\partial_t\mathbb{\Gamma}^{(2)}
-
\bbchi_{\text{P}}^{(0)}
\left(\mathbb I^{(1)}-c^{(2)}\right)
\mathbb{\Gamma}^{(2)}
\nonumber\\
&\qquad\quad+
\bbchi_{\text{P}}^{(0)}
\left(\mathbb I^{(2)}+c^{(3)}\right)
\mathbb{\Gamma}^{(1)}\mathbb{\Gamma}^{(1)}.
\end{align}
Now we use the result for  $\bbchi_P^{(0)}$ and the equilibrium identities $\mathbb{I}^{(1)}-c^{(2)}=-(\mathbb{\Gamma}^{(1)}_0)^{-1}$, and $\mathbb{I}^{(2)}+c^{(3)}=-(\mathbb{\Gamma}^{(1)}_0)^{-1}\mathbb{\Gamma}^{(2)}_0\left[(\mathbb{\Gamma}^{(1)}_0)^{-1},(\mathbb{\Gamma}^{(1)}_0)^{-1}\right]$, to write $\bbchi_P^{(1)}$ explicitly in terms of  response functions,
\begin{align}
2\bbchi_P^{(1)}\left( \bbchi_P^{(0)}\right)^{-1}\partial_t&=\bbchi_P^{(0)}\left(\mathbb{\Gamma}^{(1)}\right)^{-1}\mathbb{\Gamma}^{(2)}\left[\left(\mathbb{\Gamma}_0^{(1)}\right)^{-1},\left(\mathbb{\Gamma}_0^{(1)}\right)^{-1}\right]\nonumber\\
&-\bbchi_P^{(0)}\left(\mathbb{\Gamma}_0^{(1)}\right)^{-1}\mathbb{\Gamma}_0^{(2)}\left[\left(\mathbb{\Gamma}^{(1)}\right)^{-1},\left(\mathbb{\Gamma}^{(1)}\right)^{-1}\right].
\end{align}
Plugging  back in Eq.~\eqref{chipapp} yields $\bbchi_P^{-1}$. We provide $\bbchi_P^{-1}[\phi]\nabla\cdot \mathbf{J}=-\bbchi_P^{-1}\partial_t\phi$, as needed in Eq.~\eqref{eq:PFT},
\begin{align}
\bbchi_P^{-1}\partial_t\phi
&=\left( \left(\mathbb{\Gamma}^{(1)}\right)^{-1}-\left(\mathbb{\Gamma}_0^{(1)}\right)^{-1}\right)\phi\nonumber\\
&-\frac12\left(\mathbb{\Gamma}^{(1)}\right)^{-1}\mathbb{\Gamma}^{(2)}\left[\left(\mathbb{\Gamma}_0^{(1)}\right)^{-1}\phi,\left(\mathbb{\Gamma}_0^{(1)}\right)^{-1}\phi\right]\nonumber\\
&+\frac12\left(\mathbb{\Gamma}_0^{(1)}\right)^{-1}\mathbb{\Gamma}_0^{(2)}\left[\left(\mathbb{\Gamma}^{(1)}\right)^{-1}\phi,\left(\mathbb{\Gamma}^{(1)}\right)^{-1}\phi\right]\nonumber\\
&+O(\phi^3)\label{eq:chi1}
\end{align}
The first line is the zeroth order result, which is given in Eq.~\eqref{eq:chi-1} in Laplace space.

\section{Work done}
\label{app:work}
In this section, we give the expressions for the work done presented in Fig.~\ref{f:harmonic-chain} (c) in the main text. Using the linear order solution for $\rho_{k,t}$ in the expression for work for a periodic potential $f_{k,t}$, the general expression for work done is given by 
\begin{align}
    W_{\text{exact}}=
\frac{Nf_k^2S_k}{4}
\sum_n
\frac{g_k^{(n)}\,\Omega t_k^{(n)}}
{1+\left(\Omega t_k^{(n)}\right)^2}.
\label{eq:wexact}
\end{align}
For DDFT, on the other hand, the work done upto the same order is given as,
\begin{align}
    W_\text{DD}=
\frac{Nf_k^2S_k}{4}
\frac{\Omega t_k^{\text{DD}}}
{1+\left(\Omega t_k^{\text {DD}}\right)^2},~~\text{with }t_k^{\text{DD}}=\frac{S_k}{\nu k^2}.\label{eq:wddft}
\end{align}
The deviation $\Delta W\equiv W_{\text{exact}}-W_{\text{DD}}$ is given as 
\begin{align}
    \Delta W=\frac{Nf_k^2S_k}{4}
\left[
\sum_n
\frac{g_k^{(n)}\,\Omega t_k^{(n)}}
{1+\left(\Omega t_k^{(n)}\right)^2}
-
\frac{\Omega S_k/(\nu k^2)}
{1+\left[\Omega S_k/(\nu k^2)\right]^2}
\right].
\end{align}
where $N=\rho_0 V$. Since $\Delta W\propto N$ at fixed $f_0$, while $\Delta W/W_{\text{DD}}$
depends only on $S_{k,t}$, the error of standard DDFT is extensive. Figure~\ref{f:harmonic-chain} (c) shows Eqs.~\eqref{eq:wexact} and \eqref{eq:wddft} for the system of interacting Brownian particles. It is also worth noting that for large
$\Omega$, it can be easily shown [using Eq.~(2) in \cite{SM}] that $\Delta W=0$.

\end{document}


\title{Supplementary material:\\
Exact Volterra series for non-equilibrium mean field dynamics}

\author{Ion Santra}
\affiliation{Institut f\"ur Theoretische Physik, Universit\"at G\"ottingen }
\affiliation{Department of Physics and Astronomy, KU Leuven }
\author{Matthias Kr\"uger}
\affiliation{Institut f\"ur Theoretische Physik, Universit\"at G\"ottingen }

\maketitle

\section{General eigenmode expansion}
For overdamped Brownian particles, an expansion in eigenmodes of the
Smoluchowski operator shows that the first-order response is a sum of
exponentials~\cite{risken}, with Fourier transform
$p_k=\int dx\,p_xe^{ikx}$,
\begin{align}
\Gamma^{(1)}_{k,t}=-\rho_0S_k\sum_n\frac{g^{(n)}_k}{t^{(n)}_k}
e^{-t/t^{(n)}_k},
\label{glass:sk2}
\end{align}
with the structure factor $S_k\equiv\langle\hat\rho_k\hat\rho_{-k}\rangle/N$
\cite{hansen2013}, $t^{(n)}_k$ and $g^{(n)}_k$ non-negative, and
\begin{align}
    \sum_ng^{(n)}_k=1.\label{eq:constraint1}
\end{align}
Equivalently, $S_{k,t}/S_k=\sum_ng^{(n)}_k
e^{-t/t^{(n)}_k}$. The short-time dynamics, which is independent of
interactions~\cite{fuchs2009mode,kruger_gaussian_2017}, imposes
\begin{align}
 \sum_ng^{(n)}_k/t^{(n)}_k=\nu k^2/S_k=1/\tau_D.   
 \label{constraint2}
\end{align}
The mean relaxation time $\tau_k$ of the unperturbed fluid, introduced
in the main text via the response function, can equivalently be written
in terms of the intermediate scattering function,
\begin{align}
\tau_k=\int_0^\infty dt\,S_{k,t}/S_k=\sum_n g^{(n)}_kt^{(n)}_k,\label{taubar}
\end{align} 
with $\tau_k\geq\tau_D$, and equality only for a single exponential. With these, Eqs.~(7)
and (12), as well as standard DDFT, can be solved analytically at
linear order in $\phi$. The fitted ISF of the WCA system in the main text
is of this form.

The simplest case is a single exponential decay, i.e., $g^{(1)}_k=1$, and $g^{(n>1)}_k=0$, resembling, e.g., an ideal gas.
In this case, the two gauge choices {\it collapse}, i.e.,  $\chi^{(0)}_{\mathrm{P};k,s}=\chi^{(0)}_{\mathrm{D};k,s}=-\rho_0k^2\nu$. Furthermore, the time nonlocal direct correlation function, $C^{(2)}_{k,s}=c^{(2)}_k$, is  time local and agrees with the equilibrium form. 
 Consequently, standard DDFT yields the 
correct dynamics. 

The dynamical structure of Eqs.~(7) and (12)  becomes apparent if the sum in Eq.~\eqref{glass:sk2} contains more than one term. It is instructive to consider the case with two terms. In this case, $C^{(2)}$ of Eq.~(9) takes the form,
\begin{align}
    C^{(2)}_{k,t}&=(c^{(2)}_k-q_k)\delta_t+\frac{q_k}{\tau_k}e^{-t/\tau_k},\label{eq:C2}
    \end{align}
with $\tau_k=g^{(1)}_kt_k^{(2)}+g_k^{(2)}t_k^{(1)}$ and $
q_k=g^{(1)}_kg^{(2)}_k(t_k^{(2)}-t_k^{(1)})^2/(\rho_0 S_k\tau_k^2)
$ highlighting the dynamical information entering the time nonlocal Ornstein Zernike equation. For the PFT gauge of Eq.~(12), we find a non-Markovian dissipation kernel, which, at the leading order in $
\phi$, reads,
\begin{align}
  \frac{ 1}{\chi_{\mathrm{P}\,k,t}^{(0)}}=-\left(\frac{\delta_t}{\rho_0\nu k^2}+q_ke^{-t/\tau_k}\right).\label{eq:chi}
\end{align}
Equations~\eqref{eq:C2} and \eqref{eq:chi} illustrate that both $C^{(2)}$ and $\chi_\mathrm{P}$ take forms of typical memory kernels, e.g., employed in viscoelastic fluids \cite{Ginot2022}, consisting of a time local term and an exponential one. We thus expect them to admit simple approximations in practice, as, e.g., an expansion in exponential functions. Fig.~\ref{fig:sm} (a) for the interacting Brownian particles shows such an example where the response function requires more than two exponentials. As expected, DDFT does not correctly predict the mean field dynamics.


\section{Simulation details}
\begin{figure*}
    \centering
    \includegraphics[width=\linewidth]{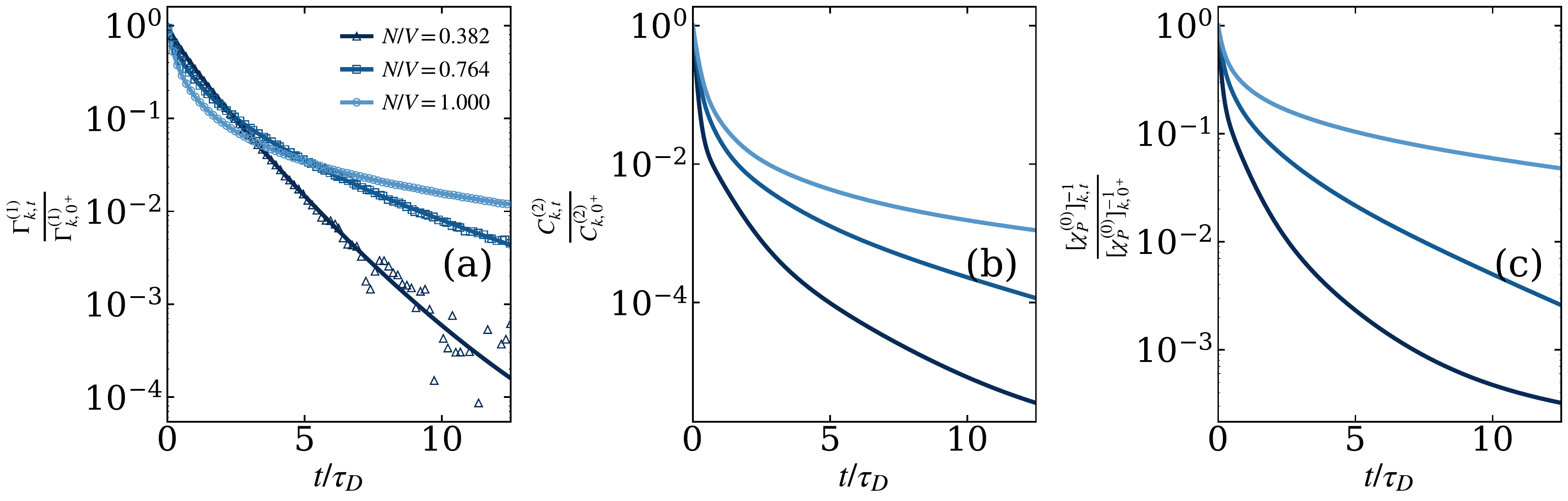}
    \caption{Components of the mean field equations for the interacting cases shown in Fig.~1 (b) in the main text. Panel (a): symbols denote the response function obtained from the numerical simulations, the solid line is the numerical fit Eq.~\eqref{glass:sk2}. (b) The time nonlocal direct correlation function obtained from Eq.~(9) in the main text. (c) The dissipation functional obtained from Eq.~(14) in the main text. In both (b) and (c) the delta-function contribution is not shown in the figure.}
    \label{fig:sm}
\end{figure*}
We simulate $N$ overdamped Brownian particles in a periodic cubic box of side
$L$ in three dimensions, using Euler--Maruyama integration of
\begin{align}
\dot{\mathbf x}_i=\nu\,\mathbf F_i+\sqrt{2D}\,\boldsymbol\xi_i,
\end{align}
with $\boldsymbol\xi_i$ Gaussian white noise of unit variance and
$D=\nu k_BT$. We use units $k_BT=\sigma=1$ and $\nu=D=1$, so that time is
measured in $\sigma^2/D$. The force is $\mathbf F_i=-\nabla_i\big[\mathcal
U_\text{int}+\sum_jf_{\mathbf x_j,t}\big]$, with
$\mathcal U_\text{int}=\sum_{i<j}u(|\mathbf x_i-\mathbf x_j|)$. We take $u$ as the
Weeks--Chandler--Andersen potential
\begin{align}
u(r)=\begin{cases}4\epsilon\big[(\sigma/r)^{12}-(\sigma/r)^{6}\big]+\epsilon,
& r<2^{1/6}\sigma,\\ 0, & r\ge2^{1/6}\sigma,\end{cases}
\end{align}
with $\epsilon=k_BT=1$. All perturbations act at the wavevector
$\mathbf k=k\,\mathbf e_x$ with $k=2\pi n/L$, i.e., at the wavevector $k$ of
the Fourier transform in the main text. The parameters are listed
in Table~\ref{tab:S_params}.

 The perturbation for panels (a) and (b) in Fig.~1 is
\begin{align}
f_{\mathbf x,t}=-f_0\cos(\mathbf k\cdot\mathbf x)\,\Theta(t)\,\Theta(t_0-t).
\label{perturbation}
\end{align}
It contains only the Fourier modes $\pm\mathbf k$.
\begin{center}
    \noindent\emph{\textbf{Estimating the kernels for the gauges from equilibrium simulations}}\\
\end{center}

We obtain the linear response from
equilibrium simulations of the unperturbed system via the
fluctuation-dissipation theorem, $\Gamma^{(1)}_{k,t}=\rho_0\,\partial_t
S_{k,t}$. The ISF $S_{k,t}$ is averaged over time origins and the three
Cartesian directions of $\mathbf k$. We fit the response by a sum of
exponentials Eq.~\eqref{glass:sk2} with the constraints Eqs.~\eqref{eq:constraint1}, \eqref{constraint2}. This is shown in for the three number densities in Fig.~\ref{fig:sm} (a). The fitted forms are then used to obtained the dynamical direct correlation function and dissipation kernel Figs.~\ref{fig:sm} (b) and (c), respectively. The static structure factor $S_k$ is measured in the same runs, and
$c^{(2)}_k=(1-1/S_k)/\rho_0$ is used for standard DDFT. Note that, we verified that the system remains a stationary fluid throughout the simulations: the intermediate scattering function $S_{k,t}$ is invariant under time translations, and the structure factor $S_k$, evaluated over successive time windows, shows no systematic change.

\begin{table}[t]
\caption{Simulation parameters for the WCA fluid
(units $k_BT=\sigma=\nu=D=1$).
The box size and wavevector are fixed while the particle
number is varied.}
\label{tab:S_params}
\begin{ruledtabular}
\begin{tabular}{lc}
Parameter & Value \\
\hline
Particle number $N$ & 500, 1000, 1309 \\
Number density & 0.382, 0.764,  1.000 \\
Box side $L$ & 10.939 \\
Interaction strength $\epsilon$ & 1 \\
Mode index $n$ & 12 \\
Wavevector $k=2\pi n/L$ & 6.893 \\
Integration time step & $2.5\times10^{-5}$ \\
Equilibration time & 150 \\
Equilibrium trajectories & 14 \\
Duration of each equilibrium trajectory & 400 \\
Independent runs (pulse) & 192 \\
Independent runs per driving frequency & 192 \\
\end{tabular}
\end{ruledtabular}
\end{table}

\noindent\emph{\textbf{Driven simulations}}\\
Each driven run starts from an independent
equilibrated configuration. We choose the amplitude such that $S_k|f_0| = 0.1$. Since the static
linear response to the field of Eq.~\eqref{perturbation} is a density modulation of
amplitude $\rho_0 S_k|f_0|$, this fixes the induced modulation to $10\%$
of $\rho_0$ at every density, small enough that the leading order in $\phi$ of Eqs.~(7) and (12) of the
main text suffices. We verified this with runs at half the amplitude,
which, rescaled by $f_0$, agree with the former within statistical error. The duration of the perturbation is chosen as $t_0 = 3\tau_k$, with the mean relaxation time of the unperturbed system $\tau_k$ given in Eq.~\eqref{taubar},
so that at each density the modulation reaches more than $90\%$ of its
stationary value before the perturbation is switched off; both the perturbed dynamics
and the subsequent free relaxation are recorded. To reduce statistical noise, each initial configuration is evolved with
$+f_0$ and $-f_0$ using the same noise realization.
Since even orders in $f_0$ do not contribute at wavevector $\mathbf{k}$
by translational invariance, half the difference is an unbiased
estimator of $\phi_{k,t}$, with strongly reduced variance.

\emph{Periodic protocol for computing work} as shown in [Fig.~1(c)]. For the WCA system, we apply
\begin{align}
f_{\mathbf x,t}=f_0\cos(\mathbf k\cdot\mathbf x)\cos(\Omega t),
\end{align}
with the same $f_0$. For each frequency, we discard a transient of at least
three periods and at least $10\,\tau_k$, and then record 20 periods. The
time step is reduced where necessary to keep at least 1000 steps per period.
The work per cycle is obtained by integrating the injected power,
$\mathcal P_t=\sum_i\partial_tf_{\mathbf x_i,t}$, over complete periods, and
averaging over periods and independent runs.

\bibliography{ref}